\documentclass[aps,prl,reprint,groupedaddress]{revtex4-1}
\usepackage{amsbsy}
\usepackage{amssymb}
\usepackage{amsmath}
\usepackage{graphicx}
\usepackage{dcolumn}
\usepackage{bm}

\begin{document}


\title{Electric field imaging using polarized neutrons}


\author{Yuan-Yu Jau}
\email[Corresponding author: ]{yjau@sandia.gov}
\affiliation{Sandia National Laboratories, Albuquerque, NM 87123, USA}
\author{Daniel S. Hussey}%
\author{Thomas R. Gentile}%
\author{Wangchun Chen}%
\affiliation{National Institute of Standards and Technology (NIST), Gaithersburg, MD 20899, USA}%


\date{\today}

\begin{abstract}
We experimentally demonstrate that electrically neutral particles, neutrons, can be used to directly visualize the electrostatic field inside a target volume that can be isolated or occupied. Electric-field images were obtained using a polychromatic, spin-polarized neutron beam with a sensitive polarimetry scheme. This work may enable new diagnostic power of the structure of electric potential, electric polarization, charge distribution, and dielectric constant by imaging spatially dependent electric fields in objects that cannot be accessed by other conventional probes.
\end{abstract}


\maketitle


Non-destructive but penetrative imaging technologies are powerful diagnostic tools, because they can reveal structure inside diagnosed objects or subjects that cannot be easily observed or measured externally.  These kinds of visualization technology have been broadly utilized in medicine, science, non-destructive testing and evaluation, security applications, etc. and it will be highly beneficial to continuously develop new imaging diagnostic technologies. One of the advantageous imaging capabilities that has been long pursued but is very challenging is the visualization of the distribution of the static electric field (${\mathbf E}$), especially when it is physically isolated and cannot be reached by physical probes. With or without the externally applied electric field, the ${\mathbf E}$ information from the diagnosed target can be used to reveal spatially dependent electrostatic characteristics, such as electric potential, electric polarization, charge distribution, and dielectric constant, which are highly valuable for studies of material properties and physical structure, examinations of electronic components and enclosed electronics, and security screening applications.

To date, the previously demonstrated and proposed approaches of static ${\mathbf E}$ imaging use physical sensors of electric potential or sensors of electric field based on field effect transistors \cite{Gebrial2006,Generazio2017}, electro-optic effect \cite{Zhu1995}, Kerr effect \cite{Zahn1994}, and Rydberg atoms \cite{Jau2020AE}. Therefore the electric field can be mapped out over the sensor locations. Although determining the electric field away from the sensor locations may be feasible in some cases, inverse problems of electric field are mostly ill-posed \cite{Kabanikhin2008}. Therefore the ${\mathbf E}$ imaging based on physical sensors is restricted to the free space around the sensor locations and the imaging resolution is limited by the distance from the imaging target to the sensors \cite{Gebrial2006,Generazio2017,Beardsmore-Rust2009}. However, this approach does not work for the occupied or physically isolated space. Hence, the previously demonstrated images are mainly ${\mathbf E}$ surveys near the surface of the diagnosed objects. In addition, owing to the technical difficulty of making a high-density, two-dimensional (2D) array of ${\mathbf E}$ sensors, a scanning scheme with a single sensor or a 1D sensor array is usually used to obtain a full image \cite{Prance1998,Gebrial2002,Generazio2017}.

To overcome the natural limitation of sensor-based ${\mathbf E}$ imaging, a novel approach is to send a probe to directly interact with the electric field, and carry the ${\mathbf E}$ information to strike the detector. From this perspective, a neutron beam would be a good candidate ${\mathbf E}$ probe, since neutrons have good penetration capability through many materials, especially metals, and the neutron spins interact with an electric field due to their motion \cite{Jau2018,Jackson1999}. In this paper, we show images of an electrostatic field for the first time using polarized neutrons. Our work may initiate a new avenue in imaging diagnostic technology.

\begin{figure*}[t]
\includegraphics[trim=20 620 0 70,angle=0,scale=1]{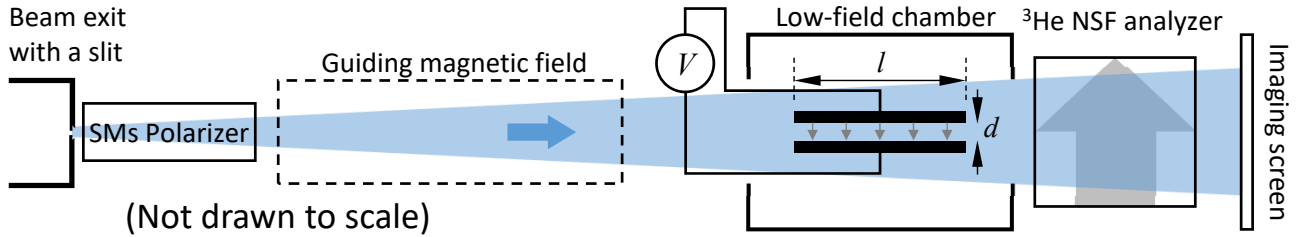}
\caption{\label{ExpSetup}A schematic of the ${\mathbf E}$ imaging experiments with polarized neutrons. The blue arrow represents the polarization of the neutron beam before arriving at the ${\mathbf E}$ region (small gray arrows), which exists in between the two parallel electrodes driven by a voltage source. The gray big arrow denotes the analyzing direction of neutron spins at the analyzer.}
\end{figure*}
The underlying physics of neutron-based ${\mathbf E}$ imaging is based on the neutron spin precession when neutrons are moving through a region of electric field. A neutron carries a non-zero magnetic moment that is aligned antiparallel to its spin orientation. The dynamics of spin precession are described by
\begin{equation}\label{NSP}
    \frac{d}{dt}\langle\mathbf{I}\rangle=\gamma_{\rm n}\langle\mathbf{I}\rangle\times\mathbf{B}_{\rm eff},
\end{equation}
where $\langle\mathbf{I}\rangle$ is the ensemble average of the neutron spin vector, $\gamma_{\rm n}=-1.83\times10^8$ rad s$^{-1}$ T$^{-1}$ is the gyromagnetic ratio of the neutron, and the effective magnetic field (${\mathbf B_{\rm eff}}$) vector seen by a moving neutron is
\begin{equation}\label{Beff}
    \mathbf{B}_{\rm eff}=\mathbf{B_{\rm lab}}-\frac{\mathbf{v}\times\mathbf{E}}{c^2},\text{ for }|\mathbf{v}|\ll c.
\end{equation}
Here, $\mathbf{B_{\rm lab}}$ and $\mathbf{E}$ are the spatially dependent  $\mathbf{B_{\rm lab}}$ and ${\mathbf E}$ vectors in the laboratory frame, $\mathbf{v}$ is the velocity vector of the moving neutron, and $c$ is the speed of light. Using Eq.~(\ref{NSP}) and Eq.~(\ref{Beff}), we find a net change in the angle vector of a neutron spin due to field-induced precession from its trajectory $l$ to be
\begin{equation}\label{DI}
    \bm{\theta}=\int_l \frac{d\langle\mathbf{I}\rangle}{|\langle\mathbf{I}\rangle|}=\int_l\gamma_{\rm n}\left[\mathbf{e}_{\rm I}\times\frac{\mathbf{B_{\rm lab}}}{|\mathbf{v}|}-\mathbf{e}_{\rm I}\times\frac{(\mathbf{e}_v\times\mathbf{E})}{c^2}\right]dl,
\end{equation}
where $\mathbf{e}_{\rm I}=\langle\mathbf{I}\rangle/|\langle\mathbf{I}\rangle|$ is the unit vector of neutron spin, and $\mathbf{e}_v=\mathbf{v}/|\mathbf{v}|$ denotes the unit vector along the velocity direction.  In practice, we can polarize neutrons to have most of their spins aligned to the same direction, and the polarization vector is $\mathbf{P}=2\langle\mathbf{I}\rangle$. The ${\mathbf  B_{\rm lab}}$ and/or ${\mathbf E}$ information then can be determined by measuring the rotation angle of the polarized neutron spins, $\bm{\theta}=\mathbf{\Delta P}/|\mathbf{P}|$ through a neutron polarimetry method. In Eq.~(\ref{DI}), we see that the ${\mathbf E}$ effect on the spin rotation is independent of neutron velocity, but the ${\mathbf  B_{\rm lab}}$ effect is velocity dependent. Hence, unlike the ${\mathbf  B_{\rm lab}}$ measurement, the ${\mathbf E}$ measurement is not affected by the velocity spread of neutrons when using a polychromatic beam. Mathematically, different neutron velocities, different probing trajectories, different polarization orientations, and different analyzing directions can be employed to retrieve full vector information of both the magnetic field and electric field in the laboratory. The experiments reported here employed cold neutrons, but if the electric field is the only interesting quantity in the measurement, in principle higher energy neutrons could be employed to minimize the effect from the magnetic field.

{$\mathbf{B_{\rm lab}}$~imaging based on neutron spin rotation signals~\cite{Kardjilov2008,Dawson2009,Strobl2009} has been demonstrated. The effects of laboratory electric fields on the neutron phase have been studied in interferometry-based Aharonov-Casher experiments~\cite{Werner2010,Cimmino1989} and the strong spontaneous electric field ($>10^9$ V/m) inside a large noncentrosymmetric crystal has been measured by neutrons~\cite{Fedorov2009}. However, to our knowledge neither ${\mathbf E}$  imaging nor measurement of spin rotation with applied ${\mathbf E}$ has been demonstrated.} One major challenge of detecting electric fields with polarized neutrons is the small magnitude of the induced spin rotation angle $\theta_{\rm E}$, which is generally much less than 1~rad due to the very small factor $1/c^2$ in Eq.~(\ref{DI}). On the other hand, the magnitude of the ${\mathbf  B_{\rm lab}}$ induced rotation angle $\theta_{\rm B}$ can easily be greater than 1~rad in many experimental conditions. Thus, in order to detect  ${\mathbf E}$ signals and produce ${\mathbf E}$ images, a neutron polarimetry method with high sensitivity of spin rotation angle is desired. In our ${\mathbf E}$ imaging experiments, we incorporated our previously developed transverse neutron polarization analysis scheme with an angular resolution $\ll$ 1 mrad ($10^{-3}$ rad) \cite{Jau2020} into a neutron imaging setup to visualize the electrostatic field produced inside experimental samples.

We conducted experiments of ${\mathbf E}$ imaging using the Neutron Guide 6 end station (NG6e)~\cite{Hussey2015} beamline at the U.S. National Institute of Standards and Technology (NIST) Center for Neutron Research (NCNR). An experimental diagram is sketched in Fig~\ref{ExpSetup}. Unpolarized polychromatic neutrons, primarily wavelengths from 0.2 nm to 0.6 nm, that left the beam exit and passed a slit were polarized by two sequential supermirror benders (SMs) to facilitate alignment. The polarized neutrons were then adiabatically transferred into a longitudinal (beam direction) guiding field of about 1 mT (10 G)). The polarized neutrons maintained their polarization and entered a low magnetic field chamber made of mu-metal with internal longitudinal ${\mathbf  B_{\rm lab}}$ strength less than 10 $\mu$T (0.1 G) and transverse ${\mathbf  B_{\rm lab}}$ strength less than 1 $\mu$T (10 mG). Inside the low-field chamber, there was a capacitor-like ${\mathbf E}$ sample with two parallel electrodes connected to a low-current, high-voltage (HV) source via HV cables. The electric field was produced in the region between the two parallel electrodes. Using Eq.~(\ref{DI}), we find the total spin precession angle through the ${\mathbf E}$ region to be
\begin{equation}\label{dth}
    \theta_E=-\gamma_n\kappa El/c^2,
\end{equation}
where $l$ is the length of the electrodes, and $\kappa$ is the correction factor that is slightly greater than one due to the fringe ${\mathbf E}$ caused by the finite length $l$. Here the electric field amplitude $E=V/d$ is defined by the driving voltage $V$ across the parallel electrodes and the spacing $d$ between the electrodes. The primary reason for placing the ${\mathbf E}$ sample inside a low-field chamber was to minimize the spin precession due to the background ${\mathbf  B_{\rm lab}}$, given the slow neutron velocity of $\approx1000$ m/s. We used a transverse polarization analysis scheme with a neutron spin filter (NSF) based on polarized $^3$He \cite{Jau2020} as shown in Fig.~\ref{ExpSetup} to provide spin-angle-dependent neutron intensity $\mathcal{I}(\theta) \propto N_0(1+\overline{P_{\rm n}A}\sin\theta)$ \cite{Jau2020} on a 300 $\mu$m thick, LiF:ZnS imaging screen. Here, $\theta$ is the projection of ${\bm\theta}$ on the analyzing direction, $N_0$ is the detected neutron number within a given area on the screen, $\overline{P_{\rm n}A}\leq1$ is the averaged product of the neutron polarization $P_{\rm n}$ and the analyzing power $A$ for different neutron wavelengths in a polychromatic beam, and $\theta=\theta_E+\theta_B$. In this experiment, $\theta_B\ll1$ can be treated as a constant background, independent of the ${\mathbf E}$ state. For $\theta\ll1$, we find the ${\mathbf E}$ signal contrast
\begin{equation}\label{Contrast}
    \frac{\mathcal{I}(\theta_E+\theta_B)-\mathcal{I}(\theta_B)}{\mathcal{I}(\theta_B)}=\overline{P_{\rm n}A}\theta_E.
\end{equation}
The LiF:ZnS screen generates scintillation photons that are proportional to the neutron intensity for optical imaging. In the ${\mathbf E}$ imaging experiments, we set the slit width to be 11 mm at the beam exit. The distances from the ${\mathbf E}$ sample to the beam exit and to the imaging screen were about 7 m and 55 cm, respectively. The vertical imaging resolution was then about 1 mm. More detailed information regarding the NG6e beamline, the experimental apparatus, and the transverse polarimetry scheme used in this work can be found in Ref.~\cite{Jau2020}.

\begin{figure}[!t]
\includegraphics[trim=30 380 0 30,angle=0,scale=0.8]{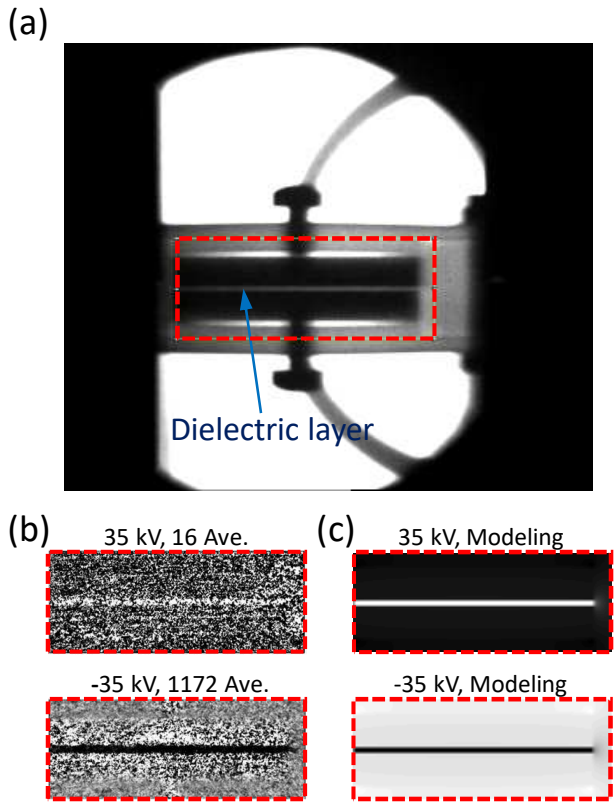}
\caption{\label{Images}(a) Normal neutron image of the short-version ${\mathbf E}$ sample. (b) Electric field images from the dashed-line selected area in (a) with two different statistics and two driving voltages of 35 kV and -35 kV on the ${\mathbf E}$ sample. (c) Modeling results of the ${\mathbf E}$ images for the same field of view. Note: for (b) and (c), the color scales and the zero-field color are different for the 35 kV and -35 kV images for the best imaging presentation. For (b), the apparent texture in the electrode region is due to higher noise produced by the strongly attenuating electrodes.  The statistics for the -35 kV image are much better than for the 35 kV image  due to much longer averaging time.}
\end{figure}
In this proof-of-concept experiment, we imaged two ${\mathbf E}$ samples (long and short versions). The body of the ${\mathbf E}$ samples are made of perfluoroalkoxy (PFA), which has high dielectric strength, and relatively high neutron transmission with low scattering. The rectangular, borated aluminum electrodes are enclosed in a  PFA body separated by a PFA membrane as the dielectric layer. In the long-version sample, the electrodes are 5 cm wide and 6.35 mm thick, $l=11.4$ cm, and the electrode spacing $d=(400\pm50)$ $\mu$m. In the short-version sample, the electrodes are 5 cm wide and 6.35 mm thick, $l=5.7$ cm, and the electrode spacing $d=(500\pm50)$ $\mu$m.  In Fig.~\ref{Images}, we compare some experimental images and the modeling results. Figure~\ref{Images}(a) is a normal neutron image of the short-version ${\mathbf E}$ sample. The two rectangular, black areas at the center are the two electrodes. The PFA dielectric layer in between the electrodes can be clearly seen. From this image, we can also see some nylon screws and the two HV cables. The usable imaging area is defined by the two straight-line boundaries on the sides and curved boundaries on the top and bottom due to the opening window on the $^3$He NSF analyzer and the opening of the low-field chamber \cite{Jau2020}. For ${\mathbf E}$ imaging, we sequentially take images with the HV source alternating between the ON and OFF states. Each image is the result of the median combination of 3 frames, with a frame exposure time of 45 s. Since the PFA material and the plastic screws in the ${\mathbf E}$ sample produced scattered neutrons that can impact the quality of the ${\mathbf E}$ image, we use a borated mask placed right before the low-field chamber to minimize the volume of the sample that is exposed to the neutron beam for reducing the scattered neutrons. The mask has a rectangular opening, and the field of view is defined by the area selected by the dashed line in Fig.~\ref{Images}(a). To produce the ${\mathbf E}$ image, we generate a contrast image by performing (IM(ON)$-$IM(OFF))/IM(OFF) from each image pair as indicated in Eq.~(\ref{Contrast}).  Here, IM(ON) and IM(OFF) denote the images with the HV source ON and OFF. We average all the contrast images to obtain better statistics. Figure~\ref{Images}(b) presents two ${\mathbf E}$ images of the short-version ${\mathbf E}$ sample at 35 kV supplied voltage with 16 averages and at -35 kV supplied voltage with 1172 averages. Since reversing the driving voltage changes the sign of the electric field and therefore the sign of  $\theta_E$, we then see the bright and dark responses on the images. Figure~\ref{Images}(c) illustrates the modeling results of the same conditions and field of view with an assumption of uniform neutron fluence. We can see qualitative agreements between the experimental and the modeling images.

\begin{figure}[!t]
\includegraphics[trim=45 0 0 40,angle=0,scale=0.34]{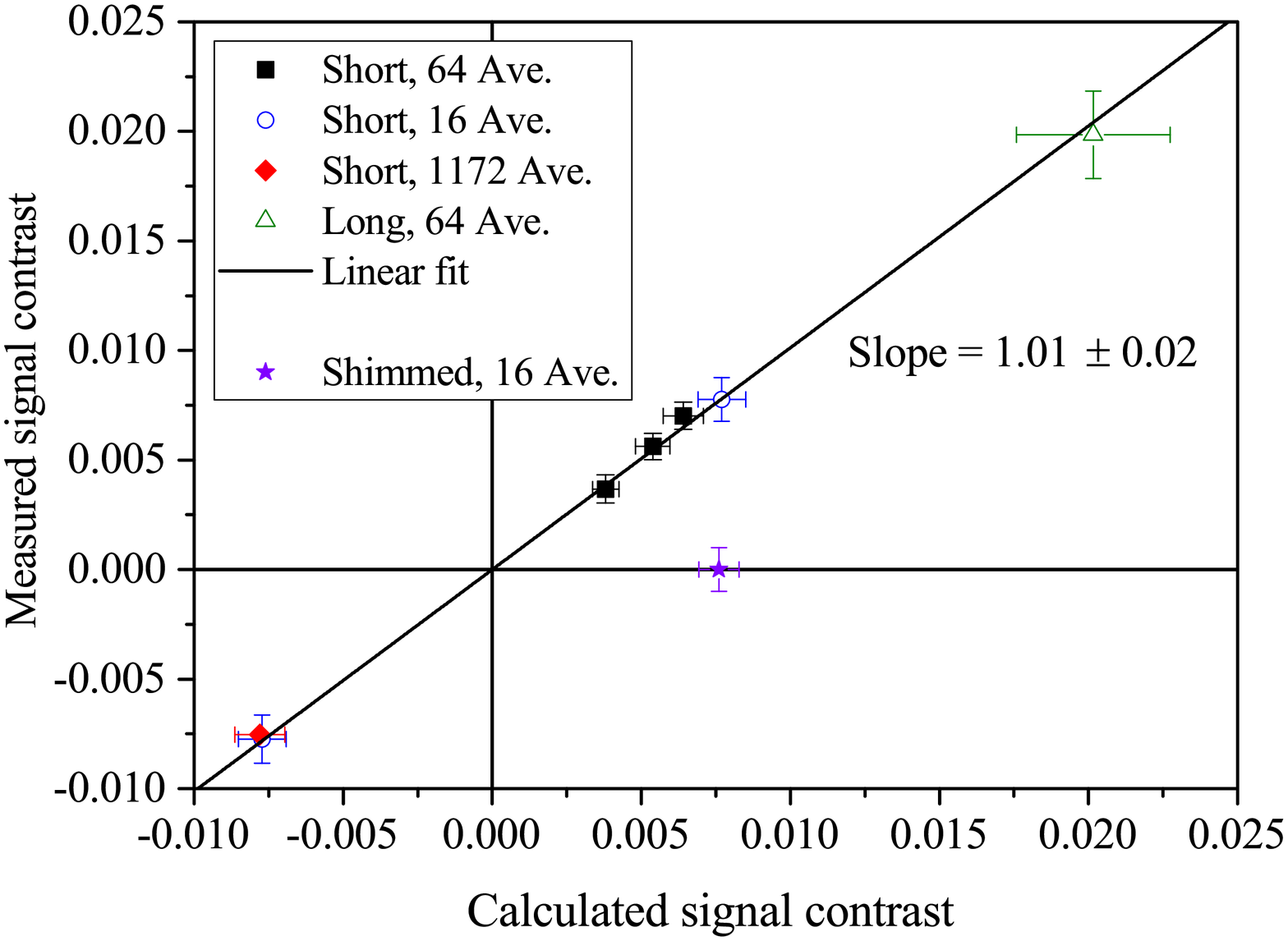}
\caption{\label{Data}${\mathbf E}$ signal contrasts vs the contrasts calculated using Eqs.~(\ref{dth}) and (\ref{Contrast}). \textquotedblleft Short\textquotedblright ~and \textquotedblleft Long\textquotedblright~represent the data points from the short-version and long-version ${\mathbf E}$ samples. \textquotedblleft Shimmed\textquotedblright~represents the data point using depolarized neutrons. The horizontal error bars are set by the machining uncertainty on the PFA membrane thickness, which is $\pm50$ $\mu$m, and the vertical error bars are limited by the imaging statistics. Also shown is a linear fit to the data, which yields a slope of $1.01\pm0.02$.}
\end{figure}
To verify the signal contrasts observed for applied electric fields, we measured the signal contrasts from the dielectric area on the images with several driving voltages from $-35$ kV to 35 kV on the short-version sample; 36kV on the long-version sample; and 35kV on the short-version sample with an iron shim (neutron depolarizer) in front of the sample. In Fig.~\ref{Data} we show the measured signal contrasts versus the values calculated using Eqs.~(\ref{dth}) and (\ref{Contrast}). From finite element modeling (FEM), we found $\kappa\approx1.032$ for the short-version sample and $\kappa\approx1.014$ for the long-version sample. The experiments were carried out over the course of a week using two different $^3$He analyzer cells that were each operated for a few days due to their exponential relaxation times of $\approx200$ h. Corrections were applied for the weak decay of the analyzing power for each cell. Based on each cell's properties and the polarizing efficiency of the supermirror pair, the range of $\overline{P_{\rm n}A}$ was between $\approx0.96$ and $\approx0.85$ over the course of the experiment, which was consistent with a direct measurement performed as discussed in Ref.~\cite{Jau2020}. The slope of a linear fit to the data yields $1.01\pm0.02$, hence the results are consistent with expectations. We see zero signal contrast with the shimmed case as a proof of the need for polarized neutrons for ${\mathbf E}$ imaging. The neutron transmission is about 10.5~\% through the short-version sample and 1.6~\% through the long-version sample. With the short-version sample, the detected neutron fluence rate on the imaging screen was about $10^3$ cm$^{-2}$ s$^{-1}$ in the dielectric area based on the gray-value readings from the image. In the experiments, the ${\mathbf E}$ image with the best statistics is the $-35$ kV driving voltage on the short-version sample with 1172 averages of the contrast images. Using the entire ${\mathbf E}$ signal area (about 5 cm by 1 mm, defined by the dielectric layer), we find the relative standard uncertainty in the measured $\theta_E$ is about 1.5~\%. The shot-noise-limit angular resolution $\delta\theta$ of the neutron spin precession can be calculated by $({P_{\rm n}A}N_0)^{-1/2}$ \cite{Jau2020,Snow2015}. We find $N_0=$ (neutron fluence rate)$\times$(area)$\times$(frame number)$\times$(exposure time)$\times$(image~number)$=10^3~\mathrm{cm}^{-2}~\mathrm{s}^{-1} \times 0.5~\mathrm{cm}^2 \times 3 \times 45~\mathrm{s} \times 1172= 7.9\times10^6$ .  Hence, $\delta\theta=0.12$ mrad. The $\theta_E$ from -35 kV on the short-version sample is $-8.4$ mrad. We then find $\delta\theta/|\theta_E|=1.4$~\%, close to the measured uncertainty. Hence, the minimum detectable voltage on the short-version sample is about 500 V using the 1172-averaged contrast images.

We estimate that for NG6e, a combination of increased sample transmission, primary aperture size, screen efficiency, and polarizing efficiency (use of a single supermirror or NSF polarizer) could allow a factor of 100 increase in neutron fluence.  In addition, beam focusing~\cite{Hussey2018,Jorba2019} could provide an additional factor of 100, yielding a total detected neutron fluence of $10^{12}$ cm$^{-2}$ in a day.  For this level, we find that the minimally detectable ${\mathbf E}$ strength for 1 cm$^3$ resolution to be about $5\times10^4$~V/m (equivalent to 500 V across 1 cm distance) and 1 mm$^3$ resolution to be about $5\times10^6$ V/m (equivalent to 5000 V across 1 mm distance). As we can see, trading sensitivity for higher volumetric resolution is inevitable. We believe this kind of ${\mathbf E}$ sensitivity is sufficient for several image diagnostics applications for targets like: high-voltage electronics, which usually contain capacitors with internal ${\mathbf E}$ strengths $>10^7$ V/m; dielectric materials with externally applied very high ${\mathbf E}$; and ferroelectric materials, which usually have spontaneous electric polarization with equivalent ${\mathbf E}$ strength $>10^8$ V/m.

We have demonstrated direct images of an electrostatic field in parallel plate capacitors using a polarized, polychromatic neutron beam. In the future, we can implement the capability of spin flipping the polarized neutron beam or flipping the analyzing direction of the NSF analyzer. This will enable ${\mathbf E}$ imaging without the need of changing the ${\mathbf E}$ state in the sample and will also enhance the signal contrast by a factor of 2 with the same experimental time. Before this work, visualizing electric field within an occupied
diagnostic space was not feasible. In addition, owing to the great penetration capability of neutrons through metals, this neutron-based ${\mathbf E}$ imaging technology can also measure the electric field that is inside a shielded space, which cannot be achieved by any other existing technology. Our work may enable new diagnostic power of the structure of electric potential, electric polarization, charge distribution, and dielectric constant inside an investigated target by visualizing spatially dependent electric field from a distance.

\begin{acknowledgments}
This work was supported by the Laboratory Directed Research and Development program at Sandia National
Laboratories. Sandia is a multimission laboratory managed and operated by National Technology and Engineering Solutions of Sandia (NTESS), LLC, a wholly owned subsidiary of Honeywell International Inc., for the U.S. Department of Energy National Nuclear Security Administration under contract DE-NA0003525. We acknowledge the support of the National Institute of Standards and Technology, US Department of Commerce, in providing the neutron facilities used in this work. The NIST effort was partially supported by the U.S. Dept. of Energy, Office of Nuclear Physics, under Interagency Agreement 89243019SSC000025. This paper describes objective technical results and analysis. Any subjective views or opinions that might be expressed in the paper do not necessarily represent the views of the U.S. Department of Energy or the United States Government.
\end{acknowledgments}

\bibliography{PNEIRefs}

\end{document}